# 六论以用户为中心的设计：智能人机交互的人因工程途径

## 许 为 *

（浙江大学 心理科学研究中心，杭州 310058）

**研究要点：**

1. 分析智能人机交互（iHCI）的人因工程特征
2. 提出"面向用户的 iHCI"的设计新理念和 iHCI 人因工程新框架
3. 采用新理念和新框架分析智能自动驾驶车人机共驾的人因问题以及协同认知生态系统，提出人因工程解决方案
4. 分析 iHCI 中的用户意图识别和人机协同合作问题，并提出今后人因工程研究的重点

**摘要** 本文从"以用户为中心的设计"理念出发，在分析智能人机交互（iHCI）的人因特征和提出的"面向用户的 iHCI"新理念的基础上，进一步提出一个基于协同认知系统、情景意识认知理论、智能代理理论的 iHCI 人因工程新框架。采用以上新理念和新框架，分析智能自动驾驶车人机共驾的人因问题以及协同认知生态系统，提出人因工程的初步解决方案。最后，分析 iHCI 中的两个基本问题（用户意图识别、人机协同合作），指出今后人因工程研究的重点。

**关键词** 以用户为中心的设计，智能人机交互，协同认知系统，人因工程，人机协同合作, 用户意图识别，人机共驾

## User Centered Design（VI）：
## Human Factors Engineering Approaches for Intelligent Human-Computer Interaction

### XU Wei

(Center for Psychological Sciences, Zhejiang University, Hangzhou 310058, China)

**Abstract**

Starting from the design philosophy of "user-centered design", this paper analyzes the human factors characteristics of intelligent human-computer interaction (iHCI) and proposes a concept of "user-oriented iHCI". The paper further proposes a new human factors framework for iHCI based on the theories of joint cognitive systems, situation awareness, and intelligent agents. With the help of the concept and framework, the paper analyzes the human factors issues in the ecosystem of autonomous vehicle co-driving and proposes an initial human factors solution. Finally, the paper analyzes the two important research areas in iHCI (i.e., user intention recognition, human-computer collaboration) and points out the key human factors questions for future research.

---

*作者: 许为, 博士,研究员；e-mail: weixu6@yahoo.com.

## 1. 引言

基于人工智能(AI)技术的各类智能系统正在改变传统人机交互(human-computer interaction，HCI)领域的人机关系模式和研究范式(许为，2005)。智能系统中的智能体（intelligent agent，或称为智能代理）在特定操作场景中具备情境感知、用户意图识别、自主学习、自主决策、自主执行等能力，由此带来了一个新的研究领域：智能人机交互（intelligent HCI，以下简称 iHCI），同时也给人因工程（human factors engineering）研究带来了新机遇（许为，葛列众，2018,2020；许为，2019a）。传统人机交互的人因工程研究主要针对人与非智能系统之间的交互，iHCI 的研究和设计需要一种新的研究范式和设计思维。 iHCI 是 AI 技术与人机交互技术不断发展和融合的产物，目前已经进入人们日常的工作和生活中。例如，语音交互，人脸识别等。在交互模式上，iHCI 利用视觉、听觉等多模态交互，提高人机交互的自然性和有效性；从交互技术来看，iHCI 正在向用户生理和情感计算、用户意图识别等方面发展；在交互理念上，iHCI 注重人机协同合作等方面，强调"以用户为中心"的人机组队（human-machine teaming）合作（O'Neill et al., 2020；Babsal et al., 2019，许为，2020）；从研究层面来看，iHCI 正在促使人、智能系统以及社会技术系统之间的紧密结合，解决智能技术的责任、伦理、道德等问题（吴亚东等，2020）。

本文从"以用户为中心的设计"理念出发（许为，2003），在分析 iHCI 的人因工程特征和提出"面向用户的 iHCI"设计新理念的基础上，进一步提出一个基于协同认知系统、情景意识认知理论以及智能代理理论的 iHCI 人因工程新框架；并且以自动驾驶车人机共驾模式为例进一步分析该框架的应用；最后，讨论和分析 iHCI 中的一些人因基本问题以及今后研究的重点。本文的目的是从人因工程的角度出发，倡导人因工程设计的新理念，提出针对 iHCI 研究和应用的人因工程新思路，推动这方面工作的开展。

## 2. 智能人机交互及人因特征

不同于传统的人机交互，iHCI 研究的对象是人类用户与基于 AI 技术的智能系统之间的交互。人因工程的总思路是通过基于"以人为中心设计"理念的一系列人与智能系统交互的研究、建模、设计、评估等活动，开发出符合用户需求、有用的、自然的、有效的、安全的 iHCI 系统。iHCI 是人-AI 交互这一新型交叉学科领域中的重要研究内容之一（许为，葛列众，高在峰，2021）。

从系统架构角度看，iHCI 主要有以下三种形态：（1）传统的非智能用户界面（系统前端）+ 智能后端系统，例如一个在线的智能化产品推荐系统或信息决策系统，它的用户界面可能是简单的图形用户界面（GUI），其智能后端系统拥有一个或多个提供智能化功能（信息推理和决策等）的智能体；（2）智能人机界面 + 非智能后端系统，例如一个传统数据库系统，它可以采用一个基于智能技术的语音交互人机界面作为用户输入，该界面具有智能交互代理的职能，建立起用户与系统后端系统之间有效交互的桥梁；（3）智能人机界面 + 智能后端系统，例如拥有智能语音交互人机界面的智能化产品推荐系统或者信息决策系统。

以上这三种 iHCI 形态的共同点就是系统都带有一个或多个智能体，用户使用这些系统时，他们与系统的智能体交互，而这些智能体拥有基于 AI 技术的一些自主化特征，在某些使用场景中表现出类似人类认知的感知、学习、自适应、独立执行等能力。正是这些智能体带来的自主化特征给 iHCI 带来了区别于传统人机交互的许多新特征。表 1 从人因工程角度概括了从传统人机交互（HCI）向智能人机交互（iHCI）过渡中所表现出的这些新特征（许为，2020；许为，葛列众，2018, 2020；许为，葛列众，高在峰，2021）。

表 1 传统人机交互（HCI）与智能人机交互（iHCI）之间的特征比较

| 过渡特征 | 传统非智能人机交互（HCI）特征 | 智能人机交互（iHCI）特征 |
| --- | --- | --- |
| 从"用户单向式"到"人机双向式"交互 | 计算系统被动地接受用户输入，只能够单向地服从于用户，接受和执行用户指令，并且根据算法和规则做出相应的输出反应 | 人与智能体的交互是双向的，智能系统可以通过感应系统来捕获和理解用户生理、认知、情感、意图等状态以及环境上下文情景等信息，主动地启动人机交互任务（例如环境智能，脑机界面）。人与智能体借助模型可预测对方行为，双向适应对方，双向均可分享信任、情景意识、意图、决策控制等 |
| 从"刺激-反应式"到"人机协同合作式"交互 | 人机之间的交互主要基于"刺激-反应"的物理关系。人机系统只有人类操作员这一单一认知体，非智能系统不拥有智能自主化等类似人类的认知特征 | 作为拥有自主化特征的机器智能体，智能体与人的交互中可以进行自主感知、理解、自主学习、自主执行等，人与智能系统（智能体）成为协同合作的队友，整个人机系统可以成为协同合作的两个认知体，分享信息、任务、目标、控制等 |
| 从"简单属性"到"情境化"交互 | 机器输入的感知目标主要是人、机、物等简单属性，例如显示器目标位置、颜色、移动轨迹等 | 智能系统输入的感知目标更具"情境化"：通过对操作场景上下文、用户行为等数据，针对"情景化"特征进行智能推演（例如用户行为体征刻画，用户消费行为画像，城市交通流量情境等），从而提供适合当前场景、满足用户需求的系统输出 |
| 从"单模态精准输入式"到"多模态模糊推理式"交互 | 基于人类的精准输入形式（通常是单一精准的输入），例如键盘、鼠标输入，系统不必关注用户行为和意图等状态，但是限制了人机交互的应用 | 智能系统有可能从不确定性条件下，从基于多模态通道的人机模糊交互（例如用户内在的交互意图，多样的应用场景，随机的交互信号数据和环境噪声）中推理出用户意图，并做出合适的系统反应，提高人机交互的自然性和有效性 |
| 从"人类智能"到"人与机器智能互补"的交互 | 机器不拥有智能自主化特征能力，人机系统中只有人类智能，不存在人与机器智能之间的互补 | 人类的生物智能(人的信息加工等能力)与机器智能(模式识别、推理等能力)之间可形成互补，在人机交互中形成更强大的、可持续发展的人机混合智能 |
| 从"用户体力负荷"到"用户认知负荷"换置 | 机器借助于自动化等技术主要替换单调、重复的人类体力工作负荷和作业任务 | 智能体还可以替换用户的认知工作负荷，其中包括主动地接管或被委派认知作业（知觉、推理、决策等） |
| 从"显式"到"隐式"交互 | 人机交互界面以"有形"显性方式呈现，例如图形化用户界面 | 人机交互界面可以"无形"的方式呈现，例如普适计算可以采用手环、穿戴设备（服装等载体）等"无形化"智能系统形式；智能环境系统可以主动启动依据用户行为、认知、生理、行为、意图状态以及特定上下文等信息的隐式人机交互 |
| 从"有限自适应"到"智能化自适应"交互 | 根据用户和环境信息输入，基于固定的计算逻辑和算法，系统有可能实现"有限的自适应"，即根据设计事先所能预测的一些操作场景产生动态化系统输出 | 自适应成为智能系统的"标配"，即没有自适应特性，就不存在智能人机交互。根据对用户、环境上下文等各种状态的感应识别、推理，智能体有可能在设计无法预测的一些操作场景中产生合适的自适应化系统输出 |

由表 1 可知，智能时代 iHCI 的这些新特征扩大了 iHCI 研究和设计的范围，远远超出了传统人机交互研究和设计的范围。因此，我们需要一种新思维和方法论来考虑如何更加有效地解决这些 iHCI 新特征所带来的新问题。另外，表 1 所列的许多新特征目前还没有完全实现，它们既反映今后 iHCI 技术发展的方向，也对今后的人因工程和 iHCI 研究提出了新要求。

## 3. 智能人机交互的设计理念

本系列文章（四论"以用户为中心的设计"：以人为中心的人工智能）提出的"以人为中心 AI"是指导 AI 系统开发的设计理念（许为，2019b），该理念同样适用于 iHCI 的研究和应用。"以人为中心 AI"理念框架包括技术、人因（human factors）、伦理三方面。本文进一步拓展"以人为中心 AI"的开发理念到 iHCI 研究和应用中，提出"面向用户的 iHCI"的设计理念(见图1)。图1 概括了围绕三方面工作的主要途径（见图1中围绕三个周边圆形部分的字体），例如用户需求和应用场景；图1 也定义了"面向用户的 iHCI"研究和应用的设计目标（见图1中围绕"面向用户的 iHCI"中心圆形部分的字体），例如自然的、可用的、有效的人机交互。

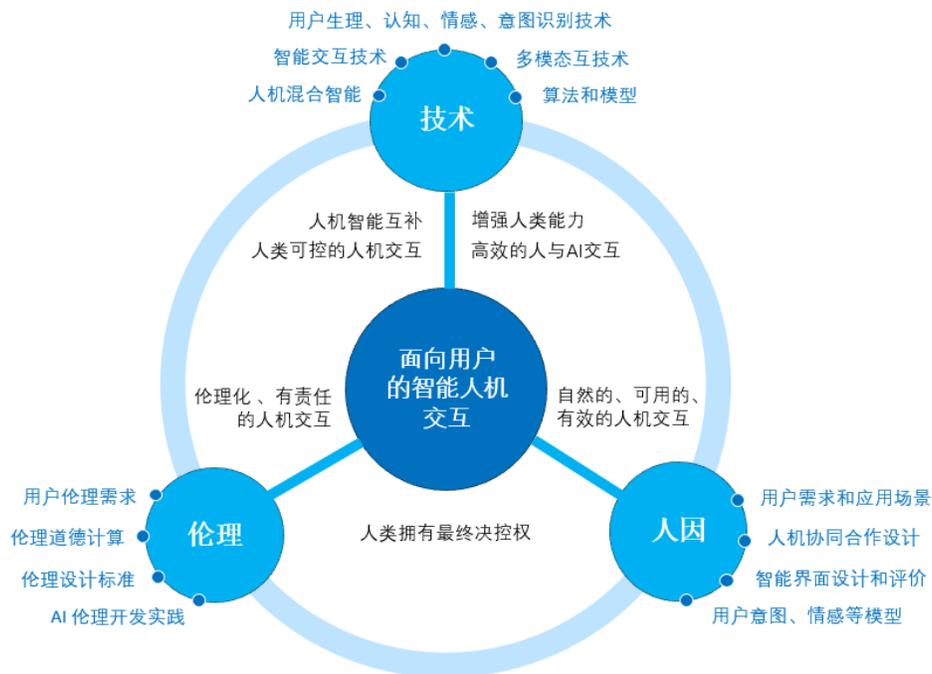

**图 1 "面向用户的智能人机交互（iHCI）"设计理念**

"面向用户的 iHCI"设计理念主要强调以下三方面的工作：

（1）"技术"方面：强调开发用户状态（如生理、认知、情感、意图等）识别技术、用户稳定特征（如人格等）识别技术、多模态交互技术、算法和模型等来支持"面向用户的 iHCI"的实现。另外，AI 界已经认识到单独发展 AI 技术的路径遇到了瓶颈效应，在高级人类认知方面难以达到人类的智能水平（Zheng et al., 2017），因此，"面向用户的 iHCI" 理念强调将人的作用融入智能人机系统,开发人机混合增强智能，达到人机智能互补。iHCI 技术开发的目的是开发出高效的、人可控制的 iHCI 系统，提升人的能力，而不是取代人。

（2）"人因"方面：强调在 iHCI 研发中从用户需求出发，落实有效的应用场景，开发有效的用户认知、意图识别、情感等模型；基于人机协同合作的设计思维，在研发中实施基于"以用户为中心"的 iHCI 设计和方法(例如用户研究、建模、用户界面原型设计、用户体验测评)。目的是开发出有用的（满足用户需求、有使用价值）、自然可用的（易用、易学）、人类拥有最终决控权的 iHCI 系统。

（3）"伦理"方面：强调从用户伦理需求（公平、隐私、道德、人的决策权等）出发，遵循伦理设计标准，通过有效的伦理道德计算、AI 伦理开发实践等方法。目的是开发出伦理化、负责任的 iHCI 系统。

进一步地，图 1 所示的"面向用户的 iHCI"设计理念体现了以下特征并且对 iHCI 研究和设计提出了以下要求：

首先，"以用户为中心"的理念。"技术"方面的工作从用户角度来设计和创新适用的交互模式，通过智能传感、智能用户意图推理等方式构建自然可用的 iHCI；根据用户需求来保证机器适应人（而不是人适应机器）；通过人机智能之间的深度整合来增强人类能力；确保将人类置于智能系统的中心。"人因"方面的工作是通过让人类成为最终决策者来确保 iHCI 是人类可控的、有效的、可用以及自然的。"伦理"方面工作旨在通过提供合乎人类道德的 iHCI 来控制和解决智能系统中的伦理问题。

其次，人因、技术和伦理三方面工作的协同合作（如图 1 中连接三个方面的线条所示）。例如，一方面，如果在设计中不考虑 AI 对人类的影响（例如 AI 的伦理道德），则 iHCI 系统是不可能实现以人为中心的设计目标，最终可能会伤害人类；另一方面，符合伦理的 iHCI 需要通过人因工程设计，借助自然有效的智能人机用户界面来以确保人类操作员能够在紧急情况下快速有效地接管对系统的控制。

最后，系统化的设计思维。将人和智能系统视为一个系统，力求在人机系统的框架内开发人机智能之间的互补。"面向用户的 iHCI"理念强调从人-机-环境系统的角度出发，保证有人类（需求、生理、心理等）、智能技术和环境（物理、文化、组织、伦理、社会等）之间的最佳匹配。因此，"面向用户的 iHCI"理念强调 iHCI 系统的开发不仅仅是一个技术项目，而是一个需要跨学科协作的系统项目。

## 4 智能人机交互的人因工程分析
### 4.1 智能人机交互的人因工程框架

在分析 iHCI 的人因特征以及定义 iHCI 设计理念和目标的基础上，本节提出一个针对 iHCI 的人因工程架构，为 iHCI 研究和应用提出人因工程解决方案。

传统人机交互理论的研究主要包括人机交互评价体系、交互认知机理、可计算的交互模型等方面。在人机交互评价体系方面，Norman（1986）提出了"以用户为中心"的人机界面评价原则，但它主要是针对鼠标键盘输入和屏幕输出的传统交互界面。在人机交互认知建模方面，人们采用人类信息处理模型来预测人类用户如何与系统进行交互，通过对人类的认知建模来设计出更高效的人机界面，例如 MHP、GOMS、SOAR、ACT-R、EPIC 等模型（详见综述：王宏安等，2020）。这些模型主要针对非智能人机交互，无法有效处理复杂场景中的智能人机交互任务，也没有考虑 iHCI 中人与智能系统之间的协同合作新型人机关系，并且机器仅仅是作为一个工具来考虑（许为，葛列众，2020）。刘烨等人（2018）后来提出的人机合作心理模型已经开始考虑到人与智能系统之间潜在的人机协同合作。

如上一篇系列文章所讨论（"五论以用户为中心的设计：从自动化到智能时代的自主化以及自动驾驶车"）（许为，2021），在非智能人机交互中，非智能系统是作为一种辅助工具（例如，自动化机器）来支持人类的操作，机器本质上是通过"刺激-反应"式的"交互"来完成对人类操作的支持(Farooq & Grudin, 2016)。这些机器依赖于由人事先固定设计的逻辑规则和算法来响应操作员的指令，通过单向、非分享的（即只有人针对机器单方向的信任、情境意识、决策控制等）、非智能互补（即只有人类的生物智能）等方式来实现人机交互。尽管人机之间也存在一定程度上的人机合作，但是作为辅助工具的机器是被动的，只有人可以启动这种有限的合作。

智能技术拥有的自主化特征带来了一种新型的人机关系：人机组队（human-machine teaming）式的合作关系，智能系统中的智能体（界面智能体或者代理）不仅仅是支持人类作业的一个工具，也可以成为与人类协同合作的团队队友，这种协同合作式是是由两者之间双向主动的、分享的、互补的、可替换的、自适应的、目标驱动的以及可预测等特征所决定的随着 AI 技术的发展，未来更多的智能系统将具备这些特征（许为，2020）。人与智能系统之间的这种新型人机关系为建立 iHCI 的人因工程框架奠定了基础。基于智能时代这种新型人机关系范式，采用协同认知系统理论（Hollnagel & Woods, 2005）、情景意识认知理论（Endsley, 1995）以及智能代理理论（Wooldridge & Jennings, 1995；王祝萍 & 张皓, 2020)，本文初步提出一个 iHCI 人因工程框架（见图 2）。

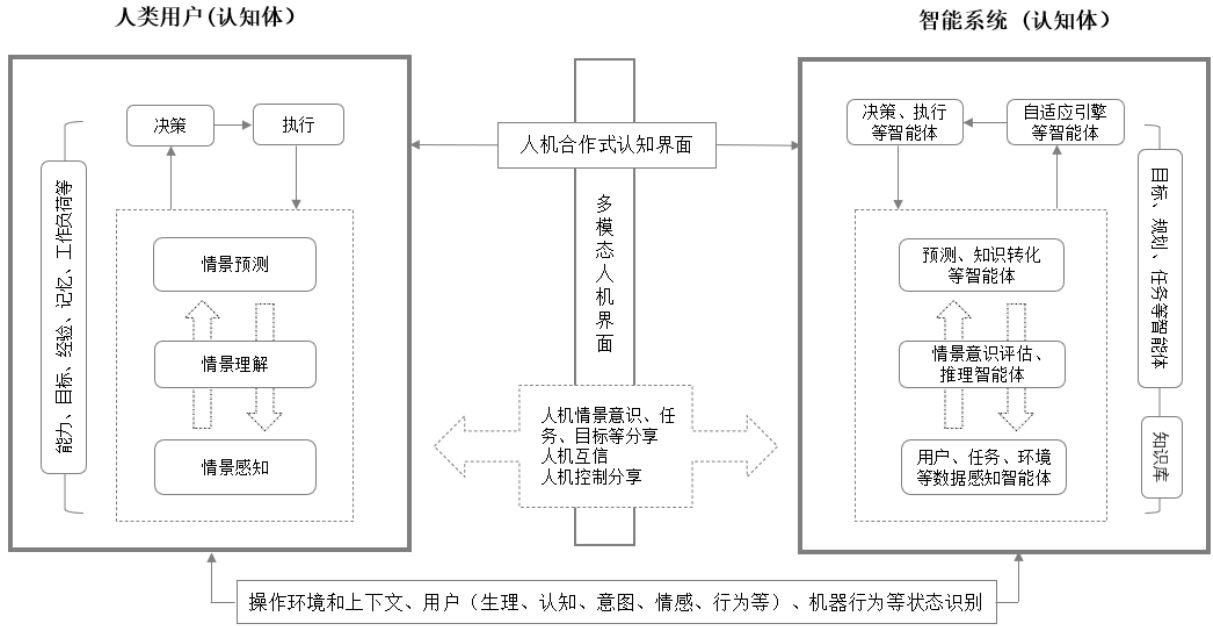

**图 2 智能人机交互(iHCI)人因工程框架的概念示意图**

相对传统人机交互框架，图 2 所示的 iHCI 人因工程框架体现了以下三方面的新特征。

第一，iHCI 人因工程框架将人类用户和智能系统（包括智能界面）均视作为能够完成一定的认知信息加工任务的认知体，从而一个 iHCI 系统可以被视为一个协同认知系统。不同于传统的人机交互系统，iHCI 是人类用户认知体与智能机器认知体之间的交互。具备自主化特征的机器智能体具有一定程度上类似人类的认知特征，可以通过自然有效的人机交互方式（如语音、手势、眼神、表情、唇动、脑电等）与人类用户开展双向的主动式交互和协同合作。在不确定性的人机交互条件下，智能体可对用户状态（认知、生理、意图、情感等状态）、环境上下文等状态进行自主感知、识别、学习、推理、理解等认知作业，并且作出自适应的自主执行。

传统的人因工程研究和应用主要关注人机界面的设计，而不是从人机协同合作的角度将人机系统作为一个整体的协同认知系统来考虑。iHCI 人因工程框架认为人和机器不是人机系统中两个独立的部分，而是一个整体，人和智能系统之间是协同合作的关系，整个认知协同系统的绩效取决于在复杂动态的操作场景中人机之间的协同合作程度，而不仅仅是某个部分的绩效。

第二，iHCI 人因工程框架采用 Endsley 的"情景意识三层次"认知理论来表征人类用户和机器认知体的智能信息加工机制（Endsley，1995，2015）。Endsley 的模型将情景意识（situation awareness，SA）定义为"人类操作员对当前环境（包括人机界面、系统、操作环境、团队合作队友等）状态的情景感知、情景理解以及对未来情景预测"的认知加工过程，该模型还包括了情景意识与记忆、经验、知识、心理模型等其他因素的认知交互。如图 2 所示（左侧），该 SA 三层情景要素（感知、理解、预测）同时拥有数据驱动（根据感知数据进行理解和预测情景）和目标驱动（根据目标以及当前的理解和预测进行基于感知数据的验证）的信息加工机制。借助于一个信息收集和后期响应的动态反馈回路机制，人类用户能够感知动态环境情景来更新获取的信息。

如图 2 所示（右侧），对应于人类操作员的信息加工机制，iHCI 人因工程框架采用与人类用户认知体异质同构性的方式来表征机器认知体的信息加工机制。该信息加工机制包括：（1）机器认知体不仅仅需要数据驱动的信息加工机制（数据感知、处理层等），同时也需要目标驱动的处理模块和反馈机制以及针对目标驱动处理的"自上而下"的决策信息加工机制；中层是对情景意识的智能评估和推理；上层则进行知

识转化，预测当前情景意识可能发生的事件；目标、规划、任务等智能体构成了反馈循环机制。（2）机器认知体拥有自适应的智能代理模块，根据环境、系统、上下文、用户状态的识别，领域知识库评估整个系统的运行情况。机器认知体基于系统任务、目标和规划，能够自适应调整系统的自主输出。（3）机器认知体拥有多智能代理，各智能代理处理相应的认知加工任务并且互相协调。

第三，iHCI 人因工程框架采用 AI 领域的智能代理理论来表征智能系统认知体的信息加工机制。智能代理是一个自治的智能计算实体，它能够不断地、自主地与环境发生交互(Wooldridge & Jennings, 1995；王祝萍 & 张皓，2020)。智能代理系统研究的目的是使智能代理能在复杂环境中实现自主运动并完成特定的任务，这需要智能代理拥有对自身状态和环境状态的建模、决策、规划和控制能力。一个理想的智能代理应该具有以下一些基本特征（Wooldridge 1995；王祝萍 & 张皓，2020）：自治性（即可以主动控制自身的行为）、目标性（对自己的意图和目标作出计划）、交互性（对环境的感知和影响）、协作性（多代理之间的协作，并且能够完成单智能体难以完成的复杂任务）、可通信性（代理之间或者代理与人之间可开展一定程度的"会话"）、学习能力、推理和规划能力、社会性、进化性、容错性、扩展性等能力。从人因工程角度看，智能代理的这些特征就是具备类似人类的认知信息加工特征，所以这些智能代理可以视为智能系统中相互交互的认知体。iHCI 人因工程框架提倡利用智能代理技术来实现"面向用户的 iHCI"的设计理念。

基于以上三方面的新特征，本文提出的 iHCI 人因工程框架能够有效地表征基于智能技术的人机交互新特征。需要指出的是，该 iHCI 人因工程框架是一个指导性的系统框架（reference architecture），它不是一个在工程上具体实现 iHCI 解决方案式的系统框架（solution architecture），iHCI 人因工程框架还有待于今后进一步研究的充实。

作为一个指导性的系统框架，针对 iHCI 的人因问题空间，iHCI 人因工程框架提出了一个基于"面向用户的 iHCI"的人因工程解决方案。表 2 进一步概括了该 iHCI 人因工程框架所表征的新特征以及人因工程解决方案的工作思路。

表 2  iHCI 人因工程框架的主要新特征以及对 iHCI 研究和应用的意义

| 新特征 | 新特征描述 | 对今后 iHCI 研究和应用的新需求 |
| --- | --- | --- |
| 协同认知系统 | 将人与智能系统表征为一个协同认知系统的两个认知体，iHCI 是人类用户认知体与机器认知体之间的协同合作 | 采用现有人-人协同合作、认知交互、认知系统工程等理论进一步研究 iHCI 理论和应用，探索 iHCI 研究和设计的新范式 |
| 人机协同合作 | 将智能时代的人机系统提升为"人机组队式合作"的新型人机关系，人类用户认知体与智能系统机器认知体可以成为合作团队队友而协同合作 | 采用人-人团队合作、协同认知系统等框架来探索人机协同合作，开发人机双向的情景意识、人机互信、人机决策和控制分享等模型（许为，2020），开发人机合作的绩效评估系统，开发社会交互、情感交互等解决方案 |
| 人机双向主动式状态识别 | 强调人机双向主动式状态识别，智能系统通过感应系统主动监测和识别用户生理、认知、行为、意图、情感等状态 | 开发有效的用户生理、认知、行为、意图、情感状态识别模型和测评方法，开发有效的知识表征和知识图谱来支持计算建模 |
| 人类智能与机器智能的互补性 | 作为一个协同认知系统，iHCI 系统的绩效不仅仅取决于系统单个部分的绩效，而且是取决于人机协同合作、人类智能与机器智能之间的有效互补。通过产生人机混合智能来最大限度地发挥人机之间的协同合作和整体系统绩效 | 开展 iHCI 中人机混合增强智能的研究和应用，为人机混合增强智能开发有效的认知计算体系架构；开发基于认知计算方法与"人在环路"方法（系统、生物学层面）整合的人机混合增强智能；优化人机混合智能的系统控制方法，探索人机融合、脑机融合等 |
| 自适应智能人机交互 | 强调智能系统的自适应机制，根据对用户、环境上下文等状态的感应识别和推理，智能 | 开发智能系统的自适应引擎代理，研究基于用户、系统、环境状态评估的自适应优化设 |

| | 体有可能在设计无法预测的一些场景中能够做出合适的自适应系统输出 | 计；利用智能系统主动式前馈预测能力来协助人类团队队友，实现主动式、自适应的人机交互 |
|---|---|---|
| 人机合作式认知界面 | 不同于传统的人机界面，人机合作式认知界面强调对人机认知体之间协同合作和认知信息加工的支持，其中包括对人机双向的情景意识、人机互信、人机决策共享、人机控制共享、人机社会交互、人机情感交互等的支持 | 开发人机合作式认知界面的设计新范式、模型和认知架构，开发针对 iHCI 的人机交互界面设计标准，提升人机界面的可用性设计，有效支持人机协同合作（例如应急场景中人机控制的快速有效交付和接管） |
| 人与智能认知体信息加工的异质同构性 | 人类操作员的认知加工模型采用 Endsley 的情景意识认知模型，该模型已经广泛应用在许多领域，并且易于测量和建模。智能系统的信息加工也采用了同样的认知架构，同时考虑智能代理的系统架构方法 | 为系统开发提供一个易于理解的认知概念系统架构（conceptual architecture），最终实现 iHCI 的协同认知系统 |

如表 2 所列，iHCI 人因工程框架对今后的 iHCI 研究和应用提出了新特征和新需求，这些新特征是传统人机交互和人因工程解决方案所不具备的，这些新需求也是目前 iHCI 研究和应用中存在的人因问题。因此，本文所提出的 iHCI 人因工程框架将有助于指导今后的 iHCI 研究和应用工作，促进"面向用户的 iHCI"设计理念和目标的实现。

### 4.2 自动驾驶车人机共驾的人因工程分析

目前，尽管人因工程、工程心理学、人机交互等专业人员参与了自动驾驶车的研发，但是频频发生的事故提醒我们需要评估目前的方法和寻找新的设计思维，从而能够进一步提升自动驾驶车的安全设计（NTSB，2017；Endsley，2018，许为，2020）。SAE（2019）将自动驾驶车系统分为 5 个"自动化"等级（L1－L5），但由于当前技术、交通法规、交通环境系统、政策法规、公众接受度等原因，完全的无人自动驾驶还有很长的路要走，在可预见的相当长一段时间内，智能自动驾驶车人机共驾将是常态，智能驾驶汽车研究和应用仍会集中在人机共驾层面（宗长富等，2021），人类驾驶员与智能自主系统仍然需要共同对车辆分时、分权地完成双方都能独立完成驾驶的任务，人类驾驶员依然扮演着重要的角色。本节采用 iHCI 人因工程框架对智能自动驾驶车的人机共驾模式做初步分析。

#### 4.2.1 单车层面分析

从自动驾驶车单车层面看，基于 AI 智能技术的自动驾驶车人机共驾模式就是一个 iHCI 系统。未来的智能自动驾驶车是一个会学习的轮式智能机器人，它与人类驾驶员一样具有一定程度的感知系统（摄像、激光雷达装备等）、智能交互界面（语音输入等）和学习等能力，在一些特定设计的场景中能够应对各种工况（马楠等，2018）。

在该车载 iHCI 中，人类驾驶员和智能车载系统是能够完成一定认知信息加工任务的两个认知体，该智能人机系统可以被视为一个协同认知系统，人机共驾体现了人类驾驶员与智能车载系统之间的交互。具备自主化特征的车载智能体具有一定程度上类似人类的认知特征，人类用户可以通过有效的 iHCI 交互方式（如语音交互）与其开展双向的交互和协同合作。装备智能感知等系统的自动驾驶车可对人类驾驶员状态、环境上下文等状态进行自主感知、识别、学习、推理、理解等认知作业，与人类驾驶员开展有效的人机共驾。

但是，人机共驾存在许多人因问题需要进一步地研究和探索，例如，有效的人机沟通模式，人机意图及情景意识和信息的双向交换，人机感知信息共享与交换，人机互信，应急状态下的人机控制权交接等等，任何一个环节的差错都可能导致安全问题（程洪等，2020；Soualmi et al., 2014）。

采用 iHCI 人因工程框架的新范式有助于开拓设计思路，探索有效的人因工程解决方案。人因工程解决方案主要包括以下几方面。

- **人机协同合作的设计新范式**：智能自动驾驶车是由两个认知体（人类驾驶员，机器认知体/"机器智能驾驶员"）组成的一个协同认知系统，两者是团队合作的队友，因此人机共驾的 iHCI 研究和应用可以从人机协同合作的新角度考虑。例如，研究要充分理解人类驾驶员与智能自主化系统在人机共驾中应该分别扮演什么角色，如何根据智能化等级和驾驶环境来确定人机共驾中人机协同合作之间的各种关系以及系统架构。基于人机协同合作的新范式，人机共驾的研究要从人机合作、人机互信、感知和情景意识共享、人机控制共享和协同驾驶等角度探索车载人机交互的优化和安全设计，从而实现有效的人机共驾(Jeong,2019；高在峰，李文敏，梁佳文等，2021)。例如，研究如何保证在应急状态下车辆控制权在人机之间的快速有效切换，确保人拥有最终控制权（包括远程控制等）(Fridman,2018)；借助人机协同合作的工作思路和有效的人机交互设计，探索在什么条件下（例如人机互信程度，人类驾驶员状态和行车意图）如何完成有效的控制权转移切换。

- **人机双向状态识别的技术**：iHCI 人因工程架构强调人机两个认知体之间双向主动的状态识别，通过机器认知体对人类驾驶员生理、认知、情感、驾驶行为和意图的监测识别和理解，以及人类驾驶员对智能系统和环境的情景意识，人机之间才能够拥有情景意识分享、人机互信、任务和目标分享、决策和控制分享等，从而有效支持人机协同合作。这方面问题需要今后的研究，例如，探索如何通过建模方法来模拟驾驶员的驾驶操作和驾驶意图；如何提高对人类驾驶员状态监测和驾驶意图识别的准确性。

- **"人在环内"的人机混合智能的系统设计**：人机共驾必须是一个"人在环内"（human-in-the-loop）的人机混合智能系统，而基于人机双向状态识别和系统控制技术为实现这样一个"人在环内"的系统安全设计提供了保障。首先，研究要从系统和 iHCI 设计角度来解决导致近几年自动驾驶车事故的人因问题（例如"人在环外"，低参与度，情景意识下降），研究在系统层面(例如"人在环内控制"，"人机协同控制")或者/和生物层面（例如脑机控制）上如何有效地实现"人在环内"的 iHCI 设计。其次，研究如何采用人机智能优势互补的设计思路来优化系统设计，提高车载协同认知系统的整体可靠性和安全性。最后，探索如何落实"有意义的人类控制"（meaningful human control）设计(Santoni & van den,2018)。例如，通过"人在环内"的 iHCI 设计，探索车载"故障追踪系统"来实现设计改善和人机故障问责的机制，推动"人类可控 AI"设计目标的实现。

- **基于智能自主化的设计思路**：SAE(2019)定义 L4-L5 等级的自动驾驶车不需要人类监控和干预，人因工程的工作思路质疑 SAE 这种设计指南。我们认为 SAE 没有基于 AI 技术的视角去评价自动驾驶车的能力，无视自动化技术和自主化技术之间的本质差别，这种对自动驾驶车智能自主化等级划分的忽略可能对设计、安全、标准化和认证产生不利影响（许为，2020）。作为一个"移动式"的智能自主化系统，高等级自动驾驶车不是传统的自动化系统，智能系统认知体拥有独特的自主化特征，这些特征对自动驾驶车 iHCI 设计提出了不同于传统自动化技术的新需求，只有充分理解这些新特征并且采取有效的 iHCI 设计来对付这些新挑战，这样才能保证自动驾驶车的行驶安全。

- **合作式认知界面设计**：基于人机协同合作的新范式，今后的研究需要探索有效的"合作式认知界面"来支持人机共驾操作中的人机协同合作、人机互信、感知和情景意识共享、人机状态识别、控制共享等问题，需要考虑有效的人机界面设计隐喻、范式以及认知架构。例如，探索在人机界面设计中如何有效地实现人机协同合作的设计方案来支持应急状态下快速有效的控制权交付和接管；探索开发有效的合作式认知界面来解决自动驾驶中的一些人因问题（例如"人在环外"，情景意识下降，自动化模式混淆）；探索基于认知工作分析（CWA）方法的车载生态用户界面（EID）的设计（Vicente, 1999; Burns & Hajdukiewicz, 2004）。

#### 4.2.2 宏观系统层面分析

从宏观社会技术系统层面来分析，安全有效的自动驾驶人机共驾的最终实现不仅仅依赖于单车层面上的有效 iHCI 设计，还需要基于网络互联、AI 与云计算、智能车联网、智能交通系统等技术；需要自动驾驶车开发企业、电信运营商、政府等各部门的合作；需要通过实现人、车、路、环境的协同交互（人与车、车与车、车与交通环境等之间）以及通过车端、路端、云端之间的信息交互，从而为自动驾驶车的安全驾驶、决策和规划提供系统化的支持。所有这些形成了一个智能自动驾驶人机共驾的生态系统（谭征宇等，2020）。

我们将 iHCI 人因工程架构所表征的人类智能体与机器智能体的协同认知系统概念拓宽到整个自动驾驶人机共驾的生态系统中，这样整个自动驾驶人机共驾生态系统可以表征为一个协同认知生态系统。该协同认知生态系统拥有一个多层次的系统架构，一个子系统的协同认知系统可以是整体协同认知生态系统架构中的某一层次。人机共驾协同认知生态系统中的各个协同认知分系统可以由 iHCI 人因工程架构所表征，例如，人类智能体与机器智能体之间（行人与智能自动驾驶车等），多重机器智能体之间（多辆自动驾驶车之间，自动驾驶车与智能道路系统之间等）。

图 3 示意了自动驾驶人机共驾协同认知生态系统的多层次架构，表 3 概括了人机共驾协同认知生态系统各层次的系统组成部分。如图 3 和表 3 所示，认知协同系统的范围和边界条件是相对的，取决于分析的目的（即系统的功能而不是它的结构）。

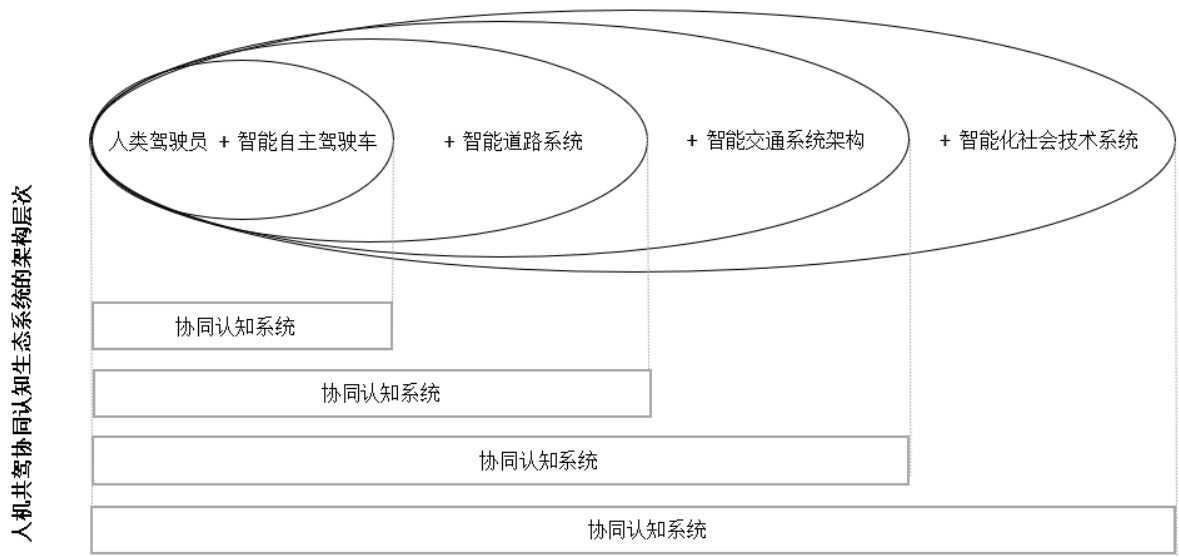

**图 3 自动驾驶人机共驾的协同认知生态系统（示意图）**

**表 3 人机共驾协同认知生态系统中各层次的主要组成部分**

| 系统组成<br>人机共驾协同认知<br>生态系统的架构层次 | 驾驶员，智能系统（环境和驾驶员状态感应系统、车载控制系统、机载智能体等） | 智能车联网,道路智能传感器，行人，其他运行自主驾驶车，道路信号，云技术，5G 等 | 智能交通信号,交通规则和法律，智能交通指挥系统等 | 公众，自动驾驶公司，电信运营商,政府等 |
|---|---|---|---|---|
| 人类驾驶员 + 智能自动驾驶车 | ✓ | | | |
| 人类驾驶员 + 智能自动驾驶车 + **智能道路系统** | ✓ | ✓ | | |
| 人类驾驶员 + 智能自动驾驶车 + 智能道路系统 + **智能交通系统** | ✓ | ✓ | ✓ | |
| 人类驾驶员 + 智能自动驾驶车 + 智能道路系统 + 智能交通系统 + **智能化社会技术系统** | ✓ | ✓ | ✓ | ✓ |

采用 iHCI 人因工程框架在宏观层面上对自动驾驶人机互驾的分析提供了以下的人因工程新思维和新途径。

首先，协同认知生态系统的宏观和系统化设计视野。自动驾驶车人机共驾的设计需要从宏观的协同认知生态系统角度来考虑，任何局限于"人类驾驶员+智能自动驾驶车"单车层面协同认知子系统的研究和设计都无法保证整个协同认知生态系统的优化设计和安全运行。

传统人因工程研究注重于人-车（"人类驾驶员 + 智能自动驾驶车"）之间的交互和人机界面设计。基于人机共驾协同认知生态系统的架构（图3），自动驾驶单车中的"人类驾驶员 + 智能自主驾驶车"仅仅是一个协同认知子系统，影响该子系统人机共驾绩效和安全的不仅仅是单车系统中两个认知体（人类驾驶员，智能自动驾驶车）之间的协同合作，还有整个人机共驾生态系统中其他层次上的协同认知系统以及相应的多重认知体。如图 3 和表 3 所示，如果将 iHCI 解决方案的分析边界逐步扩展到智能道路系统（即人-车-路系统）、智能交通系统（人-车-路-交通系统）以及智能化社会技术系统（人-车-路-交通-社会系统），实现自动驾驶车整体系统的优化设计和安全取决于各种认知体（智能体）之间的协同合作。例如，智能车联网（例如智能安全系统），道路智能传感器，路面行驶中的其他自动驾驶车（例如自动驾驶车之间的行驶状态识别、交互和协调，行车规范和行为），智能交通信号（例如信号系统智能化程度、与自动驾驶车之间的交互和兼容），智能交通指挥系统（例如系统人类操作员，决策者，智能系统），公众（例如对自动驾驶车的信任、体验），行人（例如对自动驾驶车、道路交通的认知和行为），自动驾驶车开发企业（例如人机控制权分配设计决策和设计理念），电信运营商，政府（例如智能驾驶、智能交通政策规划和决策）等等，所以这些认知体形成了一个宏观的协同认知生态系统。

由此可见，自动驾驶车的实现是一个庞大的系统工程，取决于人机共驾生态系统内所有系统组成部分之间的有效协同合作和整体系统的优化设计，低估这种系统化工程的设计思路会造成潜在的安全威胁。

其次，"以人为中心 AI"的设计理念。从"以人为中心 AI"和"面向用户的 iHCI"的理念出发，人因工程强调人机共驾模式的设计必须是"以人为中心"，应该从整个人机共驾协同认知生态系统的角度出发来保证人类拥有自动驾驶车的最终决控权。例如，在自动驾驶车人机共驾模式中，单车人类驾驶员一般拥有最终决控权。当车辆处于失控状态（例如系统故障，黑客攻击，人类驾驶员失能），一方面，车载智能系统启动智能应急方案来脱离当前的失控场景，保护人类（包括单车驾驶员、乘客、道路行人和其他车辆等），这体现"以人为中心 AI"的理念；另一方面，从系统余度化安全设计考虑，作为整个人机共驾协同

认知生态系统的分系统，智能交通系统的中央指挥控制中心操作员可远程接管故障车（例如"5G 云代驾"）来脱离当前交通场景，同时指挥协调路面其他智能驾驶车，从而保证智能道路系统的整体安全，确保"人类可控 AI"的设计目标，避免灾难性的后果。今后研究要探索如何有效地实现这种系统化设计方案。例如，在整个自动驾驶人机共驾协同认知生态系统内，研究如何根据人的能力、状态和驾驶环境进行驾驶权限和责任的实时分配，研究在应急状态下如何确保人类拥有对整个人机共驾协同认知生态系统的最终控制权限。

最后，图 3 和表 3 也示意了智能自动驾驶人机共驾生态系统发展的可能路径。这些路径包括：（1）"自下而上"的路径，即从"人类驾驶员 + 智能自动驾驶车"到"智能道路系统"、"智能交通系统"、以及"智能化社会技术系统车"；（2）"自上而下"的路径，即从"智能化社会技术系统车"到"智能交通系统"、"智能道路系统"、以及"人类驾驶员 + 智能自动驾驶车"；（3）混合或者并行的路径，即"自下而上"路径与"自上而下"路径的组合或者并行发展。混合或者并行的发展路径可能是最佳安全的路径，该结论有待进一步的研究。

## 5. 智能人机交互的基本人因问题

iHCI 中有许多人因问题，其中包括用户认知、生理计算、情感计算、用户意图识别、多模态人机交互、人机协同合作等（Xu, Dainoff et al., 2022）。限于篇幅，本节基于"面向用户的 iHCI"理念和设计目标（图 1）和 iHCI 人因工程框架（图 2），主要从人因工程角度对以下两个基本问题展开初步的讨论，并且概括了今后的一些工作。

### 5.1 用户意图识别

iHCI 中的人机双向交互需要机器智能体能够感知用户意图、操作环境和上下文，并对用户意图做出自适应的交互操作。基于用户意图识别的设计策略体现了"以用户为中心"的理念，可以降低人机冲突概率，也是实现 iHCI 中模糊和自然人机交互的关键。

用户生理、认知、情感等状态识别和计算建模为面向用户的意图识别提供了基础。用户意图识别可以通过多模态信息实现，其中包括用户生理计算（运动、视觉、听觉、脑电等），用户运动意图（表面肌电信号、脑电信号、眼电信号等），用户肢体动作（手指、手部、头动运动等），用户神经系统信号计算（如通过脑机接口技术读取用户思维活动中的交互意图信息）（张丹等，2018；程洪等，2020；Gao et al., 2014；McFarland et al., 2017）。

用户意图识别中的另一个关键是对用户意图的推理。用户意图推理主要包括两个方面：一是对用户交互意图的推理；二是对人机交互情境和上下文的推理（易鑫等，2018）。例如，在普适计算环境中，采用 Bayes 方法的用户意图推理，利用大数据用户行为、人物画像建模方法来理解以及提取用户意图。

在自动驾驶车人机共驾研究领域，人类驾驶员的意图识别和短时行为预测是重要研究内容之一，有助于提升自动驾驶车的行驶安全(宗长富等，2021)。例如，研究者使用换道意图预测系统来确定每个提示用于确定意图的相对有效性，同时利用推理分类器算法预测驾驶员的换道操作（Doshia, 2008）。验证结果表明，头部运动与车道位置和车辆动力学相结合的方法是预测换道意图的有效方法之一。基于驾驶模拟器试验数据，吕岸等（2010）结合高斯－隐马尔可夫混合模型来识别人类驾驶员在高速公路上的超车行为，并且推理分析驾驶员超车意图和行为的合理性。研究者还利用驾驶员生理信号、驾驶动态环境（例如周边车辆的运行状况）、车辆到车道边缘的距离等信息来预测驾驶员的行车意图（Tawari et al., 2014；Yuen, 2016；Mar et al., 2014；Gaikwad et al., 2015）。

iHCI 人因工程框架强调开展对用户意图识别的研究来为人机协同提供条件，从而达到自然和有效的 iHCI 设计目的。这方面有许多工作需要开展，以下列出其中的几项主要工作。

首先，目前的许多研究主要针对用户意图的识别技术,但是理论层面上有许多问题迫切需要解决,需要结合用户需求，开展人因工程研究，引导技术的发展方向。例如，用户不同通道（如动作、生理、语言等）所产生的意图需要不同的技术，从理论上如何把各类意图进行系统化地表征、建模和解释，找到多通道用户意图之间的可解释性、互补性等特征，这是人因工程的一个重要课题。另外，如何建立理解用户意图表达的认知模型和计算框架，如何有效地使用知识图谱（言语表征等)等方法来表征和理解用户交互意图(Aminer, 2020)。同时，为实现自然的 iHCI，今后的研究需要建立人与智能系统双方所需要的共同认知基础，其中包括构建与交互情境相关的知识图谱、常识性的知识推理和个性化的知识更新，帮助智能体不仅可以通过感知信息对知识图谱进行自动更新，还可以采取主动交互策略来确认和排除不可靠的推理结果(王宏安等，2020)。

其次，分析和评估用户多样性、用户行为不确定性以及使用场景多样性等因素对用户意图推理的影响(易鑫等，2018)。例如，在完成某一项动作时，由于用户可能做出不同动作所带来的不确定性，不同用户针对同一交互意图可能采用的不同交互行为，不同场景下的用户交互行为的非一致性。

第三，探索和实现从用户交互意图的模糊表达到准确识别的跨越（史元春，2019）。例如，用户交互意图是内在的，同时通过外部传感器获取的交互数据具有随机性大、背景噪声强、应用场景广泛等特点，这些问题对用户意图的准确识别和推理带来了更大的挑战。例如，在手机 iHCI 方面，史元春等人针对手机用户输入的"胖手指"问题，基于用户行为特征建模，提出了基于贝叶斯推理的用户意图理解的框架，能够从用户模糊输入信号中推测出用户的真实交互意图(Aminer,2020)。

第四，随着情感计算理论研究和技术的突破，iHCI 将向着多模态情感识别、情感意图理解、智能体多模态情感表达以及人机情感交互系统的方面深化，最终使智能系统具有主动情感认知能力(韩晶等，2015；黄宏程等，2019)。因此，加速对用户情感状态测量和建模的研究将有助于进一步开发针对用户意图识别和预测的解决方案(Wendemuth et al., 2018)。

最后，利用多模态信息的有效融合和人机交互技术来支持对用户意图的准确理解，为人机协同合作提供支持（孙效华等，2020）。虽然 AI 技术使得智能体能够比较准确地理解用户的单通道行为（例如语音识别、人脸识别），但是智能体需要借助于多通道信号的有效融合来准确理解用户的交互意图（例如用户在做什么，要做什么）（王宏安等，2020）。目前的研究主要围绕三方面探索：（1）准确识别用户已知的交互意图；(2) 准确地将已知意图的新个性化行为归纳到已知意图；(3) 准确判断和标识用户行为中出现的未知意图类别，并添加到已知学习模型。如何利用多通道信息融合方法，将已知意图的新个性化行为和新意图类别扩展到现有模型中，使得智能体在与用户的交互中学习、理解并整合新知识的能力，这将是今后的重要工作之一(杨明浩，陶建华,2018)。

### 5.2 人机协同合作

如前所述，不同于传统人机交互，iHCI 可以被视为两个认知体之间的协同合作，在一定的操作和任务环境中，人与智能体之间是基于一种团队队友之间的合作来实现两者之间的"交互"，这种交互是由两者之间双向主动的、分享的、互补的、可替换的、自适应的、目标驱动的以及可预测等人因工程特征所决定的。随着机器智能自主化程度的提高，这些特征表现更加明显（许为，葛列众，2020）。

针对人机协同合作的研究刚刚起步，许多研究从多学科角度出发，初步提出了一些解决方案。例如，人-AI 合作研究策略和框架(O'Neill et al., 2020)，人-AI 合作团队绩效评估 (Babsal et al., 2019), 人-AI 互信 (Bindewald et al., 2017),人-AI 合作中的心理模型(Bansal et al., 2019)，人-AI 合作的系统设计(Ozkaya, 2020)，人-AI 合作系统权限(Haring et al., 2019),人-AI 合作的定量和定性建模(Peters, 2019)。针对 iHCI 中人与智能体有可能成为"团队队友"的问题，Klien，Woods 等人（2004）提出了人因工程研究面临的十大挑战，其中包括人与智能体之间的可预测性（预测对方的计划、行动等），建模（表征对方的意图和行动），透明性、可解释性和理解性（针对对方意图和行为状态），双向

的协调控制等。 按照这些标准来分析，目前针对 iHCI 人机协同合作的研究和应用尚处于由"传统人机交互"向"初级人机协同合作"转变的阶段，相应的研究和应用仍然不全面，远远没有达到理想的要求。

另外，智能时代的人机交互有"双刃剑"效应。一方面，智能系统利用大数据、AI 深度学习等技术可以整合基于大量专家知识的群体智能，主动帮助人类操作员在非正常的场景中解决以往单人知识或者操作所不能解决的问题，这是传统非智能系统无法达到的；另一方面，如果 iHCI 设计中不遵循"以人为中心 AI"理念，并且不能保证人类拥有对系统的最终决策控制权，智能系统自主独立执行的结果则有可能带来伦理和安全隐患。

iHCI 人因工程框架将人机系统视为一个协同认知系统，强调开展针对人机协同合作的研究和应用，从而达到高效自然的人机交互设计目标。今后的研究应该考虑以下几个方面。

首先，要开发有效的 iHCI 中人机协同合作的理论、方法以及认知架构，进一步开发人机情景意识共享、人机互信、人机心理模型共享、人机决策共享的理论、模型和方法。例如，从技术角度进一步解决人机协同合作中的人机功能分配、动态学习和修正、主动交互模式等问题；从人机协同的体验视角进一步研究人机协同合作中的可解释性、人机互信、人机情感化、公平负责的人机协同合作等问题(孙效华等，2020)。

其次，学术界针对 iHCI 中人-智能体之间的关系有不同的理解，有待进一步研究和评估。例如，iHCI 中智能体是否真正可以成为与人类协同合作的队友（同伴、受指导者或团队领导），还是仅仅作为一个智能超级工具（Shneiderman, 2020；Klien & Woods, et al., 2004）。今后工作应该侧重于研究人类与智能体如何交互、协商以及作为队友进行协作合作；研究如何在 iHCI 设计中界定各自的角色和责任以及协作的规则；研究人机协作合作的层次结构以及各层次之间的关系（例如，协调/coordination, 协作共事/cooperation, 合作/collaboration）。从"面向用户的 iHCI"理念出发，iHCI 系统的最终策控制者应该是人，如何在人机共驾模式设计中实现这种理念需要进一步的工作。

第三，开展针对人机协同合作的计算模型和定量评估。传统非智能系统的人机交互模型（例如 GOMS，MHP）已不能满足 iHCI 中复杂交互任务的需求（例如 Card et al., 1983），需要探索人类认知机制来建模人与智能系统的交互和协作合作（刘烨等，2018），目前还没有系统化的人机交互和协作合作模型。简单明确的任务性人机交互任务相对容易建模，但是当人机交互任务包括体验性、娱乐性、沟通性以及大量的人-人之间交互活动，人机交互建模成为一项非常困难的工作，今后需要进一步开展这方面的研究。另外，人与智能体在社会交互情境中的交互也面对情境意识计算和模型的挑战，其中包括人与智能体之间的理解、意图、情绪、目标、角色结构、文化规范以及社会关系类型等，这方面的工作也有待进一步深入。

第四，开发和设计有效的人机界面来支持人机协同合作。人机两个认知体之间的协同合作必然需要能够有效支持这种人机任务的的人机界面，这是个重要的课题。例如，开发"合作式认知界面"来支持人机协同合作、人机互信、感知和情景意识共享、人机状态识别、人机控制共享等，探索基于认知工作分析的车载生态用户界面的设计（Vicente, 1999; Burns & Hajdukiewicz, 2004）。

最后，开展针对 iHCI 中长期性人机协同合作的研究。目前许多研究注重于 iHCI 中人和智能体之间的短期协同合作（例如，人与智能机器人），如何保持人和智能体之间的长期有效的协同合作仍然是一个挑战(Prada & Paiva, 2014)。例如，针对老年人、残疾人的智能康复机器人，作为认知体的机器人智能体如果无法保持与人类用户合作的兴趣，人机交互质量将大大下降。今后的研究需要解决如何能够使智能体保持与用户合适的交互和合作方式、建立关系（例如信任）、关注用户需求（例如情感需求）等问题。另外，智能体也可拥有自己的需求，也需要人类用户的帮助，今后的研究需要探索如何从 iHCI 设计上实现这种长期的人机协同合作的关系。

## 6. 总结

（1）智能技术给智能人机交互（iHCI）带来了区别于传统人机交互的许多人因工程新特征，这些新特征扩大了 iHCI 的研究范围，并且远远超出了传统人机交互的研究范围。我们需要一种新思维和方法论来考虑如何更加有效地解决 iHCI 新特征所带来的新问题，它们既反映 iHCI 技术发展的方向，也对今后的研究提出了新要求。

（2）本文进一步拓展"以人为中心 AI"的开发理念到 iHCI 研究中，提出了"面向用户的 iHCI"设计理念和目标。该理念体现"以用户为中心"的设计思想；强调人因、技术和伦理三方面的协同合作以及系统化的设计思维，从而达到"面向用户的 iHCI"设计目标。

（3）基于智能时代新型的人机协同合作关系，采用协同认知系统、情景意识认知理论和智能代理理论，本文提出一个 iHCI 人因工程框架。该框架将 iHCI 中用户和智能系统视为一个协同认知系统的两个认知体，强调"面向用户的 iHCI"设计理念和目标的实现取决于该系统中的人机协同合作、人机双向状态识别、人机智能互补、智能化自适应人机交互、合作式认知界面等方面的研究和应用。

（4）采用以上设计理念和框架，本文分别在单车层面和宏观系统层面上分析了智能自动驾驶人机共驾的人因问题，自动驾驶人机共驾提供了人因工程新思维和新途径。

（5）用户意图识别和人机协同合作是 iHCI 研究和应用中的两个基本问题，进一步开展这方面的研究将有助于实现"面向用户的 iHCI"理念和设计目标，有许多人因工程方面的问题需要进一步研究。

## 参考文献

易鑫，喻纯，史元春．(2018)．普适计算环境中用户意图推理的 Bayes 方法．*中国科学：信息科学*.

高在峰，李文敏，梁佳文，潘晗希，许为，沈模卫．(2021)．自动驾驶车中的人机信任．*心理科学进展*，29(11)，1-12.

韩晶，解仑，刘欣 等. (2015)．基于 Gross 认知重评的机器人认知情感交互模型 [J]．东南大学学报（自然科学版），45(2)，270–274.

刘烨，汪亚珉，卞玉龙，任磊，禤宇明．(2018)．面向智能时代的人机合作心理模型，*中国科学：信息科学*，48(4)，376-389.

吕岸，胡振程，陈慧．(2017)．基于高斯混合隐马尔科夫模型的高速公路超车行为辨识与分析．*汽车工程*，7，80-84.

马楠，高跃，李佳洪，李德毅.(2018)．自驾驶中的交互认知．*中国科学：信息科学*，48(8)，1083-1096.

谭征宇，戴宁一，张瑞佛，戴柯颖．(2020)．智能网联汽车人机交互研究现状及展望，*计算机集成制造系统*，26(10)．

孙效华，张义文，秦觉晓，李璟璐，王舒超. (2020)．人机智能协同研究综述．*包装工程*，41(18)，1-11.

许为. (2003)．以用户为中心设计：人机工效学的机遇和挑战．*人类工效学*，9(4)，8-11.

许为. (2005)．人-计算机交互作用研究和应用新思路的探讨．*人类工效学*，11(4)，37-40.

许为. (2017)．再论以用户为中心的设计：新挑战和新机遇．*人类工效学*，23(1)，82-86.

许为. (2019a)．三论以用户为中心的设计：智能时代的用户体验和创新设计．*应用心理学*，25(1)，3-17.

许为. (2019b)．四论以用户为中心的设计：以人为中心的人工智能．*应用心理学*，25(4)，291-305.

许为. (2020)．五论以用户为中心的设计：从自动化到智能时代的自主化以及自动驾驶车．*应用心理学*，26(2)，108-128.

许为，葛列众.(2018)．人因学的新取向．*心理科学进展*，26(9)，1521-1534.

许为，葛列众.(2020)．智能时代的工程心理学．*心理科学进展*，28(9)，1409-1425.

许为，葛列众，高在峰. (2021) 人-AI 交互：实现"以人为中心 AI"理念的跨学科新领域．*智能系统学

报, 16(4), 604-621.

杨明浩, 陶建华. (2018). 多通道人机交互信息融合的智能方法. *中国科学: 信息科学*, 48, 433-448,

吴亚东等. (2020). 人机交互技术及应用, *机械工业出版社*.

王宏安等. (2020). *人工智能：智能人机交互*, 电子工业出版社.

王祝萍, 张皓. (2020). *自主智能体系统*, 人民邮电出版社.

宗长富, 代昌华, 张东. (2021). 智能汽车的人机共驾技术研究现状和发展趋势. *中国公路学报*, *34*(6), 214.

张丹, 刘永进, 赵国朕. (2018). 自然交互的生理计算, *中国计算机学会通讯*, 14(5), 36-40.

黄宏程, 刘宁, 胡敏等. (2019). 基于博弈的机器人认知情感交互模型. *电子与信息学报*, 41(10), 2471-2478.

Aminer. (2020). 人工智能之人机交互。https://www.aminer.cn/research_report/5ed9bc15a296c11af9ae8460

Bansal, G., Nushi, B., Kamar, E., Lasecki, W. S., Weld, D. S., & Horvitz, E. (2019, October). Beyond accuracy: The role of mental models in human-AI team performance. In *Proceedings of the AAAI Conference on Human Computation and Crowdsourcing,* 7(1), 2-11.

Bass, T. (1999, May). Multisensor data fusion for next generation distributed intrusion detection systems. In *Proceedings of the IRIS National Symposium on Sensor and Data Fusion* (Vol. 24, No. 28, pp. 24-27). COAST Laboratory, Purdue University, l.

Bindewald, J. M., Rusnock, C. F., & Miller, M. E. (2017, July). Measuring human trust behavior in human-machine teams. In *International Conference on Applied Human Factors and Ergonomics* (pp. 47-58). Springer, Cham.

Burns, C. M. & Hajdukiewicz, J. (2004). *Ecological Interface Design*. CRC Press.

Card, S. K., Moran, T. P., & Newell, A. (1983). *The Psychology of Human-Computer Interaction.* Hillsdale: Lawrence Erlbaum Associates.

Deroo, M., & Hoc, J. M. (2014). Analysis of human-machine cooperation when driving with different degrees of haptic shared control. *IEEE transactions on haptics*, *7*(3), 324-333.

Doshi, A., & Trivedi, M. (2008, June). A comparative exploration of eye gaze and head motion cues for lane change intent prediction. In *2008 IEEE Intelligent Vehicles Symposium* (pp. 49-54). IEEE.

Endsley, M. R. (1995). Toward a theory of situation awareness in dynamic systems. *Human Factors,* 37, 32–64.

Endsley, M. R. (2015), Situation awareness misconceptions and misunderstandings. *Journal of Cognitive Engineering and Decision Making*, 9(1): 4-32.

Endsley, M. R. (2018). Situation awareness in future autonomous vehicles: Beware of the Unexpected. *Proceedings of the 20th Congress of the International Ergonomics Association (IEA 2018)*，*IEA 2018*, published by Springer.

Farooq, U., & Grudin, J. (2016). Human computer integration. *Interactions, 23*, 27−32.

Fridman L. (2018). Human-Centered Autonomous Vehicle Systems: Principles of Effective Shared Autonomy. MIT HCAV Research Program: https://arxiv.org/pdf/1810.01835.pdf.

Gao S, Wang Y, Gao X, et al. (2014). Visual and auditory braincomputer interfaces. IEEE transactions on bio-medical engineering, 61(5),1436-47.

Gaikwad, V., & Lokhande, S. (2014). Lane departure identification for advanced driver assistance. *IEEE Transactions on Intelligent Transportation Systems*, *16*(2), 910-918.

Haring, K. S., Mosley, A., Pruznick, S., Fleming, J., Satterfield, K., de Visser, E. J., ... & Funke, G. (2019, July). Robot authority in human-machine teams: effects of human-like appearance on compliance. In *International Conference on Human-Computer Interaction* (pp. 63-78). Springer, Cham.

Hollnagel, E., & Woods, D. D. (2005). *Joint cognitive systems: Foundations of cognitive systems engineering.* London: CRC Press.

Jeong, K. A. (2019). Human-system cooperation in automated driving. *International Journal of Human-Computer Interaction*, 35(11), 917-918.

Klien, G., Woods, D. D., Bradshaw, J. M., Hoffman, R. R., & Feltovich, P. J. (2004). Ten challenges for making automation a" team player" in joint human-agent activity. *IEEE Intelligent Systems*, *19*(6), 91-95. Engineering and Human–Machine Teaming. *IEEE Software*, *37*(6), 3-6.

McFarland D J, Wolpaw J. R. (2017). EEG-Based BrainComputer Interfaces. *Current Opinion in Biomedical Engineering*, 4(C), 194-200

Mars, F., Deroo, M., & Hoc, J. M. (2014). Analysis of human-machine cooperation when driving with different degrees of haptic shared control. *IEEE transactions on haptics*, *7*(3), 324-333. Ozkaya, I. (2020). The Behavioral Science of Software Norman, D. A. (1986). *User centered system design: New perspectives on human-computer interaction*. CRC Press.

NTSB. (2017). Collision between a car operating with automated vehicle control systems and a tractor-semitrailor truck near Williston, Florida, May 7, 2016. *Accidents Report*, by National Transportation Safety Board (NTSB) 2017, Washington, DC.

O'Neill, T., McNeese, N., Barron, A., & Schelble, B. (2020). Human–autonomy teaming: A review and analysis of the empirical literature. *Human Factors*, 0018720820960865.

Peters, N. (2019). Interruption timing prediction via prosodic task boundary model for human-machine Teaming. *In Future Information and Communication Conference*, 501-522, Springer, Cham.

Prada, R., & Paiva, A. (2014). Human-agent interaction: Challenges for bringing humans and agents together. In *Proc. of the 3rd Int. Workshop on Human-Agent Interaction Design and Models (HAIDM 2014)* at the 13th Int. Conf. on Agent and Multi-Agent Systems (AAMAS 2014) (pp. 1-10).

SAE (Society of Automotive Engineers). (2019). Taxonomy and definitions for terms related to driving automation systems for on-road motor vehicles. *Recommended Practice J3016* (revised 2019-01-07）.

Santoni de Sio, F., & van den Hoven, J. (2018). Meaningful human control over autonomous systems: a philosophical account. *Front Robot AI* , 5, 15.

Soualmi, B., Sentouh, C., Popieul, J. C., & Debernard, S. (2014). Automation-driver cooperative driving in presence of undetected obstacles. *Control engineering practice*, *24*, 106-119.

Shneiderman, B. (2020). Design lessons from AI's two grand goals: Human emulation and useful applications, *IEEE Transactions on Technology and Society, 1*, 73-82.

Vicente, K. J. (1999). *Cognitive Work Analysis: Toward Safe, Productive, and Healthy Computer-Based Work.* Hillsdale, NJ: Erlbaum.

Wendemuth, A., Böck, R., Nürnberger, A., Al-Hamadi, A., Brechmann, A., & Ohl, F. W. (2018, October). Intention-based anticipatory interactive systems. In *2018 IEEE International Conference on Systems, Man, and Cybernetics (SMC)* (pp. 2583-2588). IEEE.

Wooldridge, M., & Jennings, N. R., (1995). Intelligent agent: theory and practice, *Knowledge engineering*, 10(2).

Yuen, K., Martin, S., & Trivedi, M. M. (2016, December). On looking at faces in an automobile: Issues, algorithms and evaluation on naturalistic driving dataset. In *2016 23rd International Conference on Pattern Recognition (ICPR)* (pp. 2777-2782). IEEE. Mars, F.,

Xu, W., Dainoff, M., Ge, L., & Gao, Z. (2022, accepted). Transitioning to human interaction with AI systems: New New challenges and opportunities for HCI professionals to enable human-centered AI. *International Journal of Human-Computer Interaction*

Zheng, N. N., Liu, Z. Y., Ren, P. J., Ma, Y. Q., Chen, S. T., Yu, S. Y., ... & Wang, F. Y. (2017). Hybrid-augmented intelligence: collaboration and cognition. *Frontiers of Information Technology & Electronic Engineering*, *18*(2), 153-179.